\documentclass[10pt]{iopart}

\RequirePackage[normalem]{ulem} 
\RequirePackage{color}\definecolor{RED}{rgb}{1,0,0}\definecolor{BLUE}{rgb}{0,0,1} 

\newcommand{\equ}[1]{eq.~(\ref{equ:#1})}

\newcommand{\Fig}[1]{fig.~\ref{fig:#1}}

\usepackage{textcomp}
\usepackage{graphicx}
\usepackage{subfig}
\usepackage[usenames,dvipsnames]{xcolor}
\usepackage{url}

\usepackage{citesort}

\begin{document}


\title[]{The effect of the stochasticity of photoionization on 3D streamer simulations}

\author{B. Bagheri$^1$, J. Teunissen$^{1,2}$} 

\address{$^1$Centrum Wiskunde \& Informatica, Amsterdam, The Netherlands\\$^2$Centre for Mathematical Plasma-Astrophysics, KU Leuven, Belgium}
\ead{b.bagheri@cwi.nl}
\vspace{10pt}
\begin{indented}
\item[]\today
\end{indented}

\begin{abstract}
  Positive streamer discharges require a source of free electrons ahead of them
  for their growth. In air, these electrons are typically provided by
  photoionization. Here we investigate how stochastic fluctuations due to the
  discreteness of ionizing photons affect positive streamers in air. We simulate
  positive streamers between two planar electrodes with a 3D plasma fluid model,
  using both a stochastic and a continuum method for photoionization. With
  stochastic photoionization, fluctuations are visible in the streamer's
  direction, maximal electric field, velocity, and electron density. The
  streamers do not branch, and we find good agreement between the averaged
  stochastic results and the results with continuum photoionization. The
  streamers stay roughly axisymmetric, and we show that results obtained with an
  axisymmetric model indeed agree well with the 3D results. However, we find
  that positive streamers are sensitive to the amount of photoionization. When
  the amount of photoionization is doubled, there is even better agreement
  between the stochastic and continuum results, but with half the amount of
  photoionization, stochastic fluctuations become more important and streamer
  branching starts to occur.
\end{abstract}

\ioptwocol

\section{Introduction}
\label{sec:introduction}

Streamers~\cite{vitello_simulation_1994,yi_experimental_2002,ebert_review_2010}
are rapidly growing electric discharges with an elongated shape. They can appear
when the electric field in a nonconducting medium exceeds the breakdown
threshold. Streamer channels are surrounded by a space charge layer that
enhances the electric field at their tips, where electron impact ionization
causes them to grow. Due to their electric field enhancement, streamers can
propagate into regions where the background field is below the breakdown
threshold. Streamers typically occur in gases~\cite{nijdam_probing_2010},
although they can also form in liquids~\cite{An_2007}.

Streamers are the precursors to lightning leaders and sparks, and they can be
observed directly as sprites~\cite{sentman_red_1995} above thunderstorms. In
technology, streamers are used in diverse
applications~\cite{Fridman_2005,Adamovich_2017}, for example for the production of
chemical radicals~\cite{kanazawa_observation_2011}, in ignition and
combustion~\cite{starikovskaia_plasma-assisted_2014} and plasma catalysis
\cite{nozaki_non-thermal_2013}. Because streamer channels are weakly ionized,
they typically do not significantly increase the gas temperature.

Streamers can be of positive or negative polarity. Positive streamers propagate 
typically in the direction of the electric field. Negative streamers propagate in the opposite direction. In this paper we
focus on positive streamers, which in air form more easily than negative ones
\cite{briels_positive_2008}. Because positive streamers propagate against the
electron drift velocity, they require a source of free electrons ahead of them.
When these electrons enter the high-field region around a positive streamer head
they rapidly multiply due to electron-impact ionization, which causes the
streamer channel to extend. The propagation of positive streamers can therefore
be quite sensitive to the distribution of free electrons ahead of
them~\cite{nijdam_role_2016}.

In air, photoionization is typically the most important source of free electrons
\cite{nijdam_probing_2010,pancheshnyi_role_2005} ahead of positive streamers.
This process occurs when a UV photon emitted by an excited nitrogen molecule
ionizes an oxygen molecule. A commonly used model for photoionization is
Zheleznyak's model \cite{zheleznyak_photoionization_1982}. Recent work has
suggested several refinements to Zheleznyak's model, such as the generalization
to other gas mixtures \cite{pancheshnyi_photoionization_2014}, the inclusion of
multiple excited states and transitions~\cite{Stephens_2016,Stephens_2018} and
taking the lifetime of excited states into account~\cite{Jiang_2018}.

Numerical solutions to Zheleznyak's photoionization model are often computed
using the so-called Helmholtz approximation, in which the
absorption function of the produced photons is written in terms of an
exponential expansion~\cite{luque_photoionization_2007,bourdon_efficient_2007}.
The photoionization profile can then be computed as a density from a set of Helmholtz
equations.
In reality, photoionization is a stochastic process, in which the ionizing
photons are discrete. Discrete photoionization events can be generated using
Monte Carlo methods, as was first done in~\cite{chanrion_pic-mcc_2008}.

Stochastic fluctuations due to photoionization are likely to dominate over
fluctuations due to e.g., the discreteness of electrons, because of the
relatively low photon numbers and long absorption distances. The goal of the
present paper is therefore to study how fluctuations due to photoionization
affect positive streamers in atmospheric air. We investigate the role of these
fluctuations using 3D plasma fluid simulations, and compare results with
stochastic photoionization to results with a continuum approximation for
photoionization. We also study how sensitive the evolution of positive streamers
is to the amount of photoionization. Furthermore, the 3D results are compared to
axisymmetric simulations performed with a model that was recently benchmarked
against five other codes~\cite{bagheri_comparison_2018}.

\emph{Comparison with earlier work:} In~\cite{Xiong_2014}, a fluid model was
used to study the effect of stochastic photoionization on streamer propagation
and branching in a 2D Cartesian
geometry. 
An important difference with the present paper is that in this work we perform
3D simulations, which are required to realistically capture stochastic
fluctuations. The role of stochastic electron density fluctuations for streamer
branching was investigated in~\cite{Luque_2011}, and in
\cite{li_comparison_2012} the effect of such fluctuations on the onset of
branching for negative streamers was studied, but in overvolted gaps and without
photoionization. There exist several papers about PIC (particle-in-cell)
simulations of streamers including stochastic photoionization, for
example~\cite{chanrion_pic-mcc_2008,Teunissen_2016,Stephens_2018}, but these
simulations are typically limited to short propagation lengths due to their high
computational cost. Another new contribution of this paper is that we compare
stochastic photoionization to the common continuum approach.

The structure of the paper is as follows. The plasma fluid model and simulation
conditions are described in section \ref{sec:discharge-model}, after which the
photoionization approaches are described in section \ref{sec:photoi-model}. In
section \ref{sec:results} stochastic and continuum photoionization are compared
using both 3D and axisymmetric simulations. Furthermore, the number of UV
photons is varied to show when stochastic effects become important.

\section{Discharge model and conditions}
\label{sec:discharge-model}

We use a plasma fluid model of the drift-diffusion-reaction type coupled to
the local field approximation, as implemented in
\cite{teunissen_simulating_2017}. A fluid model cannot correctly capture all the
physical noise in a simulation, but on the other hand, PIC
simulations typically introduce too much noise, when they employ
super-particles. Using a fluid model for the electron density allows us to single out the
noise due to the discreteness of the photons and the photoionization. 
Another advantage of fluid models is that they are
computationally much cheaper than PIC codes.

In \cite{bagheri_comparison_2018}, our plasma fluid model
\cite{teunissen_simulating_2017} for streamer discharges was compared with
models from five other groups. Axisymmetric simulations of positive streamers
were performed. For sufficiently fine grids and small time steps, reasonably
good agreement between the models was found.

The comparison study~\cite{bagheri_comparison_2018} included one test case with
photoionization, using the Helmholtz approximation. Here we generalize that test
case in two ways: First, stochastic fluctuations due to photoionization are
taken into account. Second, the simulations are performed in 3D, to
realistically model fluctuations and their effect on the streamer's propagation.
We use the same fluid model, transport coefficients, computational domain and
photoionization parameters as in~\cite{bagheri_comparison_2018}.

\subsection{Model equations}
The electron density $n_e$ and positive ion density $n_i$ evolve in time as
\begin{eqnarray}
  \partial_t n_e &= \nabla \cdot (n_e\mu_e\mathbf{E} + D_e\nabla n_e)
                   + S_i + S_\mathrm{ph},\\
  \partial_t n_i &= S_i + S_\mathrm{ph},
  \label{equ:fluid-model}
\end{eqnarray}
in which $\mu_e$ is the (positive) electron mobility, $D_e$ the electron
diffusion coefficient, $\bar{\alpha}$ the effective ionization coefficient,
$\mathbf{E}$ the electric field, $S_i = \bar{ \alpha}\mu_e |\mathbf{E}|n_e$ the
ionization source term and $S_\mathrm{ph}$ the non-local photoionization source
term (see section \ref{sec:photoi-model}). Ion motion is neglected. The electric
field is computed in the electrostatic approximation as
\begin{eqnarray*}
  \mathbf{E} &= -\nabla \phi,\\
  \nabla^{2}\phi &= -\frac{e(n_i-n_e)}{\epsilon_0},
\end{eqnarray*}
where $\phi$ is the electric potential, $\epsilon_0$ the vacuum permittivity,
and $e$ the elementary charge.

We consider streamer discharges in dry air, containing $80\%$ $\textnormal{N}_2$
and $20\%$ $\textnormal{O}_2$, at $p=1$~bar and $T=300$~Kelvin. The local field
approximation is used for the transport coefficients, so that $\bar{\alpha}$,
$\mu_e$ and $D_e$ depend on the local electric field strength. We use the same
analytic transport coefficients as in \cite{bagheri_comparison_2018}, which were
retrieved from \cite{dutton_survey_1975,hartmann_theoretical_1984}. At a gas
pressure of $p=1 \, \textrm{bar}$ and a gas temperature of
$T=300 \, \textrm{K}$, the coefficients can be written as
\begin{eqnarray*}
  \mu_e &= 2.3987 \, E^{-0.26}\\
  D_e &= 4.3628 \times 10^{-3} \, E^{0.22}\\
  \bar{\alpha} &= \alpha - \eta\\
  \alpha &= (1.1944 \times 10^6 + 4.3666 \times 10^{26}/E^3)
                    e^{-2.73 \times 10^7/E} \\
  \eta &=340.75,
\end{eqnarray*}
where SI units have been omitted, so that $E$ is the electric field strength in
$\mathrm{V/m}$, $\mu_e$ the mobility in $\mathrm{m}^2/(\mathrm{V \, s})$ etc.
More accurate transport coefficients could be obtained by using a Boltzmann solver~\cite{hagelaar_solving_2005}, but for the purpose of this study having highly realistic transport coefficients is not essential.

The fluid model used here is described in more detail
in~\cite{teunissen_simulating_2017}. It is based on the Afivo
framework~\cite{Teunissen_afivo_2018}, which contains geometric multigrid
methods to quickly solve Poisson's equation, octree-based adaptive mesh
refinement and OpenMP parallelism. The fluid equations are solved using explicit
second order time stepping, and a slope-limited second order accurate spatial
discretization.

\subsection{Computational domain and initial conditions}
\label{sec:comp-domain}

\begin{figure}
  \centering
  {\includegraphics[width=0.4\textwidth]{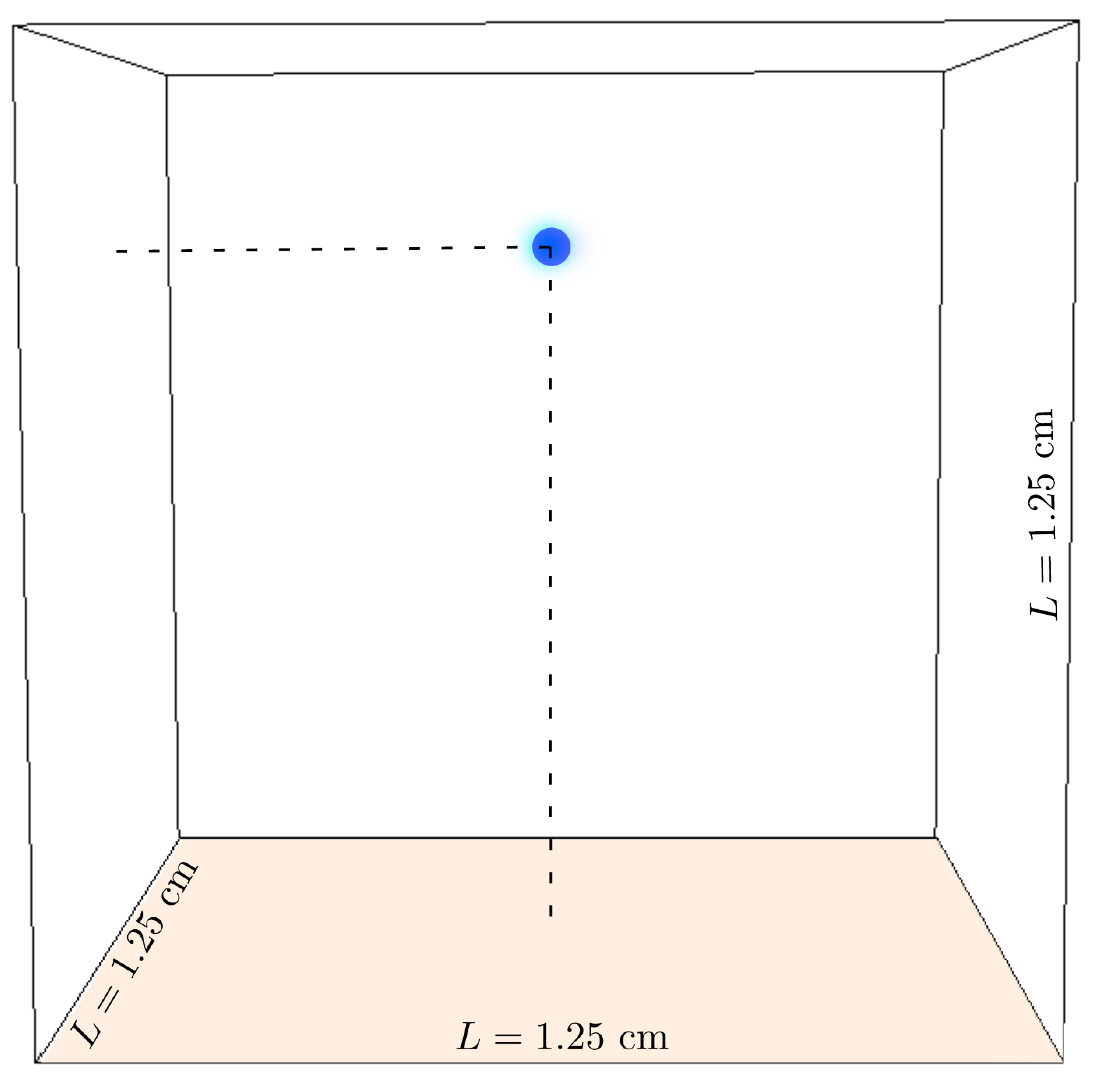}}
  \caption{A 3D view of the computational domain showing the position of the
    initial seed from which the streamer starts. The seed has a Gaussian
    distribution and a maximal density of $5\times 10^{18}~\textnormal{m}^{-3}$
    positive ions, a width of 0.4~mm, and it is located at a height of 1~cm. The
    top plane is at a potential of $18.75$~kV and the bottom plane (colored) is
    grounded.}
  \label{fig:computational-domain}
\end{figure}

The computational domain shown in \Fig{computational-domain} is used for 3D
simulations. It measures 1.25~cm in all three Cartesian directions. The top
plane is at a potential $\phi=18.75$~kV and the bottom plane is grounded; hence
the background field is 15~kV/cm, which is about half of the breakdown field.
For the potential, homogeneous Neumann boundary conditions are used on the other
four sides, and for the electron density they are used on all boundaries.
Instead of a needle electrode protruding from the top plate, a positive immobile
seed with a Gaussian distribution is implemented. The seed is located at a
height of $1$~cm, and it has a maximal density of
$N_0=5\times 10^{18}~\textnormal{m}^{-3}$ positive ions, and a width of
$\sigma=0.4$~mm:
\begin{equation}
  n_i(\mathbf{r})=N_0\exp{ \left[-\frac{(\mathbf{r}-\mathbf{r}_0)^2}{\sigma^2} \right] },
  \label{equ:gaussian-seed}
\end{equation}
where $\bf{r}_0$ indicates the location of the seed. A homogeneous background
density of $n_e=n_i=10^9 \, \textnormal{m}^{-3}$ electrons and positive ions is
included. This background density facilitates the start of the discharge, after
which photoionization will quickly dominate over it\footnote{As pointed out by
  one of the referees, a fluid model cannot accurate simulate the (stochastic)
  first electron avalanches originating from such a low background density. Here
  we have chosen to only focus on the stochastic effects of photoionization, not
  on the stochasticity of discharge inception.}.
For the axisymmetric simulations, we use the same computational domain as in
\cite{bagheri_comparison_2018}.

We use the same refinement criterion as in \cite{teunissen_simulating_2017},
namely refine if $\alpha(1.2 \times E) \, \Delta x > 1.0$, where $\alpha(E)$ is
the field-dependent ionization coefficient, $E$ is the electric field strength
in V/m, and $\Delta x$ is the grid spacing. For the simulations in
section~\ref{sec:results} this gives an AMR grid with a minimum grid spacing of
about $3 \, \mu\textrm{m}$.

\section{Photoionization models}
\label{sec:photoi-model}

In this study, we make use of Zheleznyak's photoionization model for air
\cite{zheleznyak_photoionization_1982}. Assuming that ionizing photons do not
scatter and that their direction is isotropically distributed, the photoionization source term $S_{\mathrm{ph}}(\mathbf{r})$ is given by
\begin{equation}
S_{\mathrm{ph}}({\bf r})=\int d^3r'\;\frac{I({\bf r'})f(|{\bf r}-{\bf r'}|)}{4\pi|{\bf r}-{\bf r'}|^2},
\label{equ:photo-general}
\end{equation}
where $I(\mathbf{r})$ is the source of ionizing photons,
$4\pi|\mathbf{r}-\mathbf{r}'|^2$ is a geometric factor, and $f(\mathbf{r})$ is
the absorption function that gives the probability density of photon absorption
at a distance $r$:
 \begin{equation}
f(r)=\frac{\exp(-\chi_{\mathrm{min}}p_{O_2}r)-\exp({-\chi_{\mathrm{max}}}p_{O_2}r)}{r\ln(\chi_{\mathrm{max}}/\chi_{\mathrm{min}})},
\label{equ:absorption-function}
\end{equation}
where $\chi_{\mathrm{max}}\approx1.5\times10^2/(\textnormal{mm bar})$,
$\chi_{\mathrm{min}}\approx2.6/(\textnormal{mm bar})$, and $p_{O_2}$ is the
partial pressure of oxygen. In Zheleznyak's model the UV photon source term
$I(\mathbf{r})$ is proportional to the electron impact ionization source term
$S_i$
\begin{equation}
I(\mathbf{r})=\frac{p_q}{p+p_q} \xi \, S_i,
\label{equ:Photo-source-term}
\end{equation}
where the factor $p_q/(p+p_q)$ accounts for the collisional quenching of excited
nitrogen molecules, where $p$ is the gas pressure and we use a quenching
pressure of $p_q = 40 \, \mathrm{mbar}$. The proportionality factor $\xi$ is in
principle field-dependent \cite{zheleznyak_photoionization_1982}, but for
simplicity we here set it to $\xi = 0.075$ (except for section
\ref{sec:Photo-electron reduction}, where it is varied).

As stated in the introduction, there have recently been several efforts to
improve upon Zheleznyak's model, for example by taking different excited states
and their lifetime into account~\cite{Stephens_2016} or by considering different
gas mixtures~\cite{pancheshnyi_photoionization_2014}. Since in this paper we focus on the
effect of stochastic fluctuations, we use Zheleznyak's model in its standard
formulation.

\subsection{Continuum (Helmholtz) approach}
\label{sec:photoi-continuum-approach}

Directly evaluating the integral in \equ{photo-general} is computationally too
expensive, in particular in 3D.
In~\cite{luque_photoionization_2007,bourdon_efficient_2007} an approximation was
proposed in which the absorption function $f$ of \equ{absorption-function} is
expanded as
\begin{equation}
  \label{eq:exp-expansion}
  f(r) \approx r \sum_{i=1}^{N} c_i \, e^{-\lambda_i r},
\end{equation}
where $c_i$ and $\lambda_i$ are fitted coefficients. When this expansion is plugged into equation (\ref{equ:photo-general}), one obtains $N$ Helmholtz
equations that can be solved with fast elliptic solvers to obtain
$S_{\mathrm{ph}}$. We remark that the exponential expansion of equation
(\ref{eq:exp-expansion}) differs by a factor of $r^2$ from the form of equation
(\ref{equ:absorption-function}). Therefore, the expansions can only be accurate
in a certain range and not for very short absorption distances. As discussed
in~\cite{luque_photoionization_2007} this is often acceptable, since photons
that travel only a short distance are usually not important for the discharge
dynamics.

In this work we consider three possible expansions for the absorption function, which we
denote by
\begin{itemize}
  \item [B2:] Bourdon's two-term expansion
  \item [B3:] Bourdon's three-term expansion
  \item [L:] Luque's two-term expansion
\end{itemize}
Luque's parameters~\cite{luque_photoionization_2007} are defined in such a way
that the $\xi$ from equation (\ref{equ:Photo-source-term}) is incorporated in
the $c_i$ coefficients from equation (\ref{eq:exp-expansion}). Therefore results
obtained with these parameters do not precisely correspond to $\xi = 0.075$,
which is used for the other photoionization methods. For details about these
differences and the parameters that are used in this paper see Appendix A of
\cite{bagheri_comparison_2018}.

%
%

\subsection{Stochastic (Monte Carlo) approach}
\label{sec:photoi-stochastic-approach}

Another approach is to use Monte Carlo methods to model photoionization as a
stochastic process, which it of course also is in real discharges due to
the discreteness of photons. This approach was
first described in \cite{chanrion_pic-mcc_2008}, where it was used for PIC
simulations. We use the implementation described in chapter 11
of~\cite{teunissen_3d_2015}, which was also used in \cite{nijdam_role_2016}. The
same approximations are made as for the continuum approach: photon scattering is
neglected and photon directions are isotropic. In the limit of an infinite
number of infinitesimal photons, the stochastic photoionization profile therefore agrees with
the solution of equation (\ref{equ:photo-general}). The computation of
stochastic photoionization profile consists of several steps:
\begin{enumerate}
  \item The discrete number of ionizing photons in each cell within a given time
  step $\Delta t$ is sampled from a Poisson distribution, with the mean given by
  \equ{Photo-source-term}. For a detailed description, see chapter 11 of
  \cite{teunissen_3d_2015}.
  \item For each photon, an absorption length is sampled from
  \equ{absorption-function}, and a direction from an isotropic distribution is determined using
  random numbers. Together, these numbers determine the absorption location of
  the UV photons.
  \item The absorption locations of the photons are mapped to grid densities
  using bi/trilinear interpolation. For this study, this mapping is always done
  on the finest available grid.
\end{enumerate}

We make use of photons with a weight $w = 1$, so that each computational
photon corresponds to a physical photon. By using \emph{super-photons} for which
$w > 1$, the computational cost of the method can be reduced, but unphysical
noise is introduced. Conversely, \emph{sub-photons} for which $w < 1$ could also
be used to reduce the physical noise in the solution.

Even when using physical photons the noise in our simulations is still somewhat underestimated, 
since we are using a fluid approximation for electrons and ions. Electrons produced in
photoionization events do not move (and ionize neutrals) as discrete particles,
but instead they correspond to advecting patches of increasing electron density. Furthermore,
when photoionization events occur in regions with a coarse grid, the fine-scale
noise in the photoionization profile is not captured.


We remark that it is also possible to sample discrete photons from a
photoionization profile computed with a continuum method. However, a pure Monte
Carlo approach has several advantages: it does not need to introduce any
assumptions about the absorption function, and it can easily be adapted to
include surface and object interactions.


\section{Results and discussion}
\label{sec:results}

\subsection{Stochastic vs. continuum photoionization in 3D}
\label{sec:3D}

\begin{figure}%
  \begin{center}
    \includegraphics[width=\linewidth]{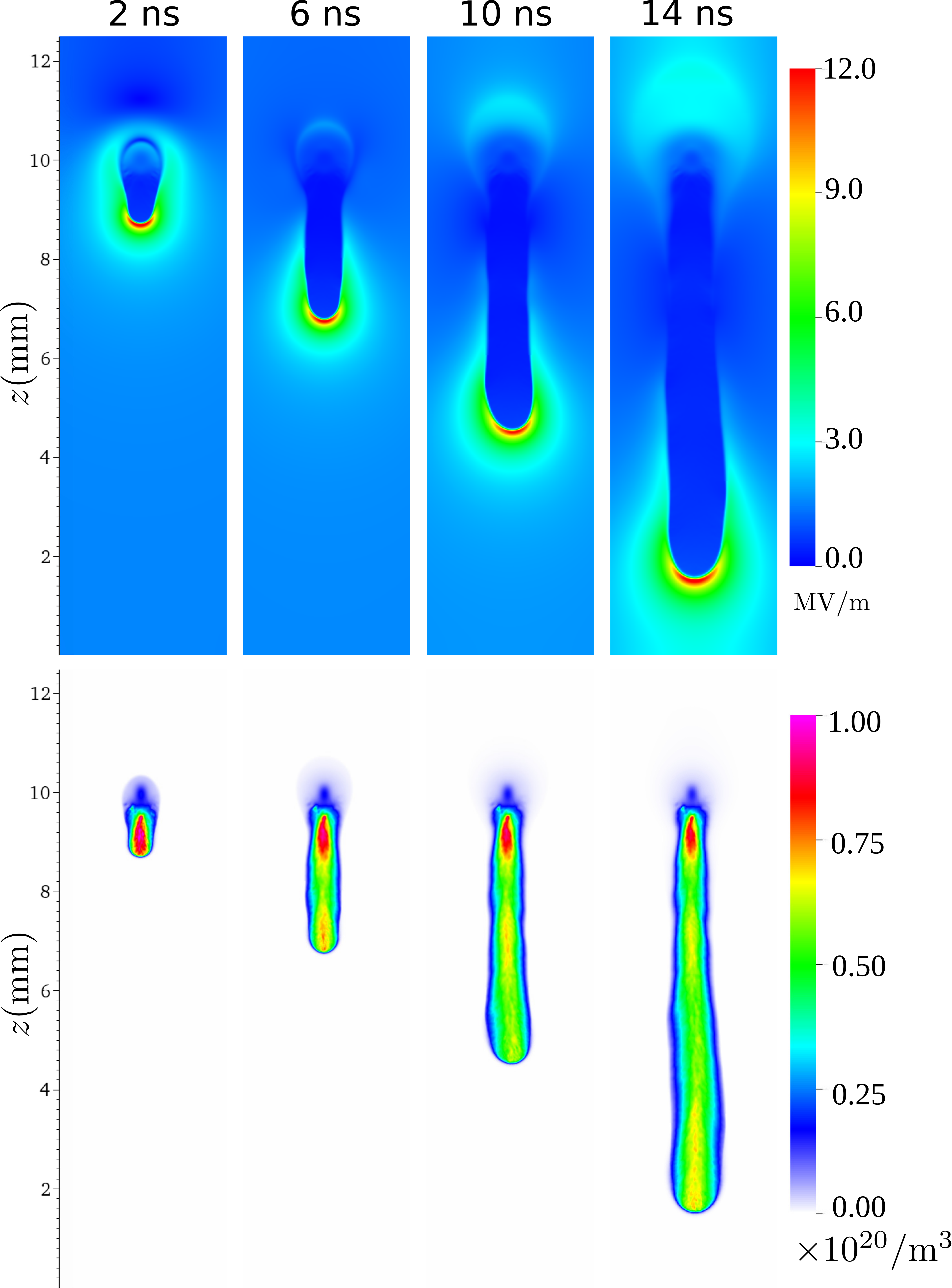}
    \caption{Cross sections in the $xz$ plane showing the evolution of the
      electric field~(top) and the electron density~(bottom) in a 3D simulation.
      Due to stochastic photoionization, fluctuations are visible in the
      positive streamer's downwards propagation. The width of the domain in the
      $x$-direction is $1.25$~cm; only a part of it is shown here.}
    \label{fig:E-ne-time-3D}%
  \end{center}
\end{figure}

In this section we compare 3D simulations with stochastic and continuum
photoionization, using the computational domain and the initial conditions
described in section \ref{sec:comp-domain}. Figure \ref{fig:E-ne-time-3D} shows
an example of the discharge evolution with stochastic photoionization.
Initially, the positively charged seed enhances the background field, and a
single positive streamer starts to grow downwards. As the streamer propagates
downwards, several small horizontal deviations are visible. After about
$16 \, \textrm{ns}$, the streamer reaches the bottom electrode.

\begin{figure*}
  \begin{center}
    \includegraphics[width=1.0\linewidth]{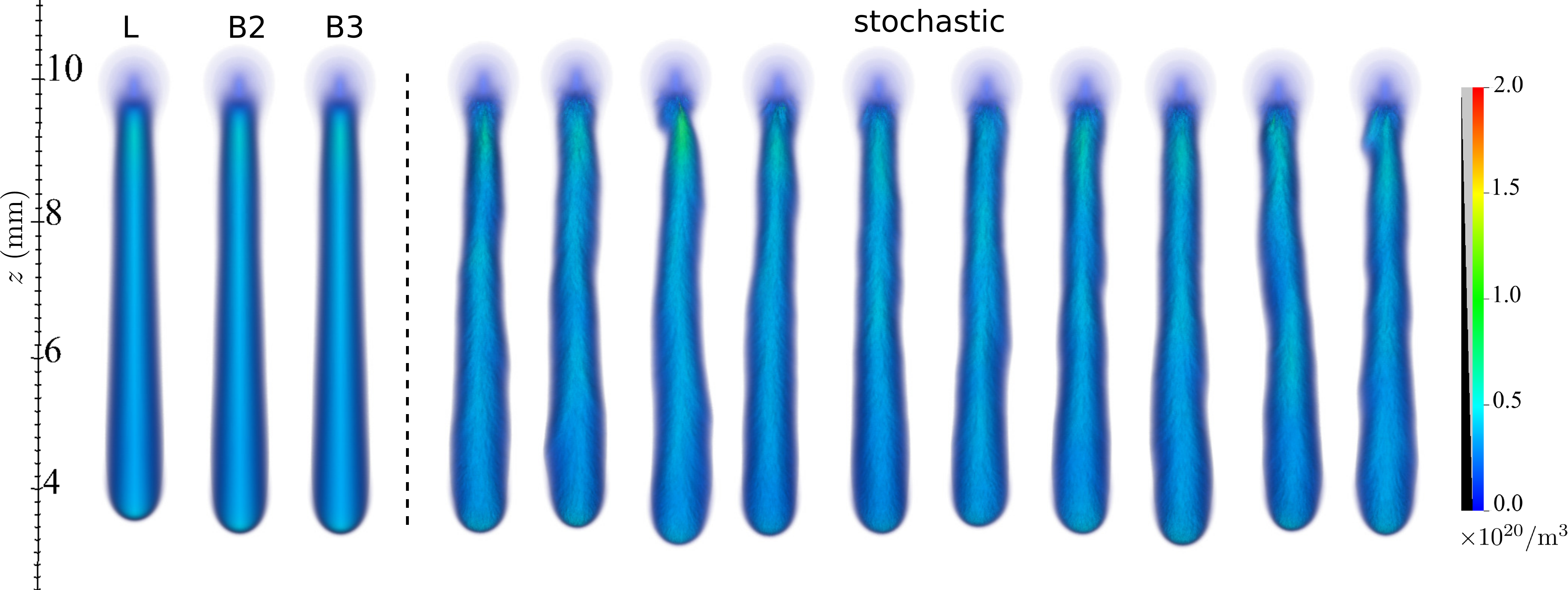}
    \caption{Volume renders of electron density at $t=13$~ns for the simulations in 3D. Ten stochastic runs, each with a different initial state of the random number generator, are shown on the right. The results with continuum photoionization (with L, B$_2$, and B$_3$ parameters) are shown on the left. }
    \label{fig:3D-diff-rnds-constdx-10um-E-plots-3}
  \end{center}
\end{figure*}

To determine the average behavior and the variability of the results with
stochastic photoionization, we performed ten runs, each with a different initial
state of the random number generator.
Figure~\ref{fig:3D-diff-rnds-constdx-10um-E-plots-3} shows volume renders of the
electron density at $t=13$~ns for these runs. For comparison, results with
continuum photoionization are also shown, using Luque's two-term and Bourdon's
two- and three-term parameters, see section \ref{sec:photoi-continuum-approach}.
In all cases the streamer bridges the gap without branching, and in terms of
propagation length there is quite good agreement between the photoionization
methods. With stochastic photoionization, small fluctuations in the electron
density and propagation length are visible.

\begin{figure}%
  \begin{center}
    \includegraphics[width=\linewidth]{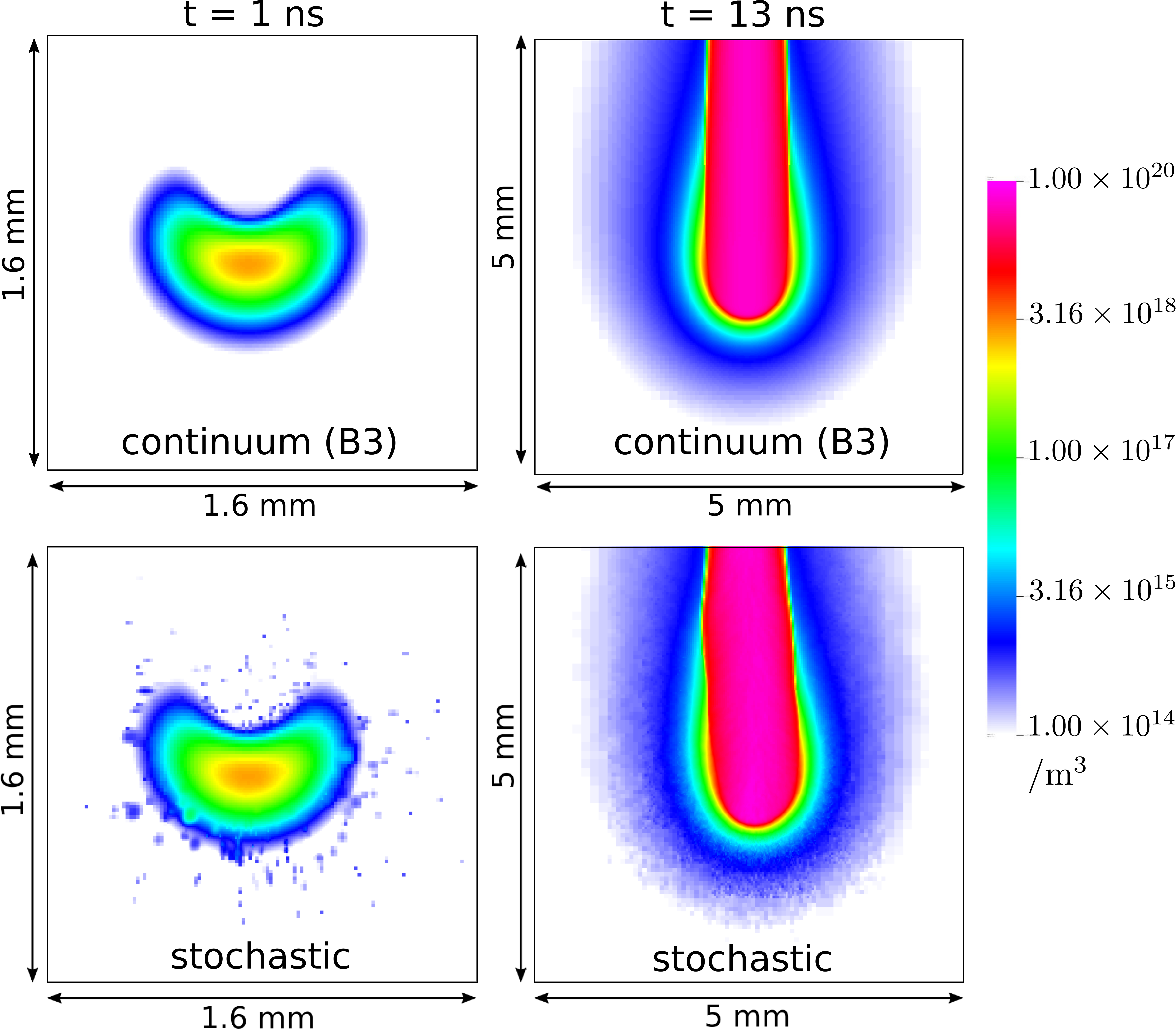}
    \caption{Cross sections of electron density in xz plane around the streamer head for the 3D simulations at $t=1$~ns and $t=13$~ns. Top: the results with continuum (B$_3$ parameters) photoionization. Bottom: the results with stochastic photoionization.}
    \label{fig:3D-diff-rnds-constdx-10um-photo-5}%
  \end{center}
\end{figure}

The stochastic fluctuations appear to be more pronounced at the beginning of the
streamer channel. The reason is probably that the noise in the photo-electron
density is larger during the start of the streamer. Later on, the region ahead
of the streamer has been exposed to photoionization for a longer period, and
more photons are produced per unit time. This is illustrated in figure
\ref{fig:3D-diff-rnds-constdx-10um-photo-5}, which shows cross sections of the
electron density around the streamer head at $t =1\, \textrm{ns}$ and
$t = 13 \, \textrm{ns}$ on a logarithmic scale. With continuum photoionization,
the electron density ahead of the streamer is smooth, whereas clear fluctuations
in the electron density are visible with stochastic photoionization, in
particular at $t =1\, \textrm{ns}$.

\begin{figure}%
  \begin{center}
    \includegraphics[width=\linewidth]{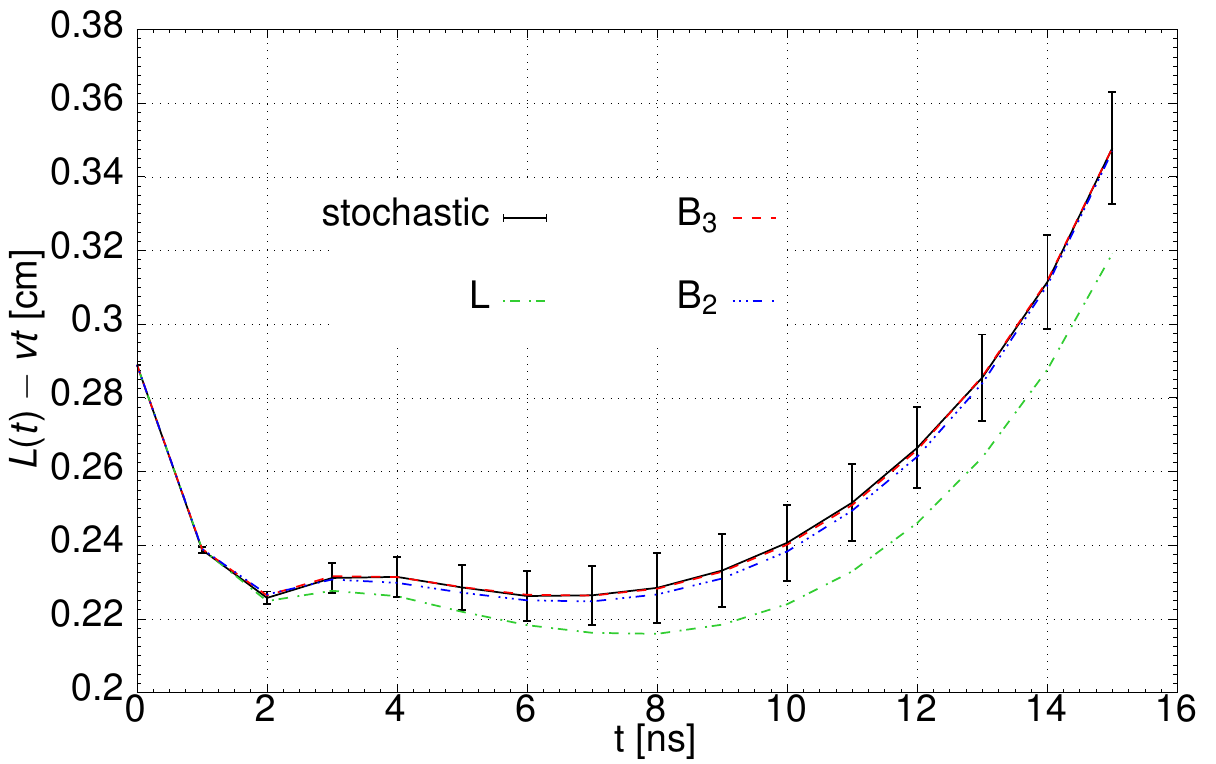}
    \includegraphics[width=\linewidth]{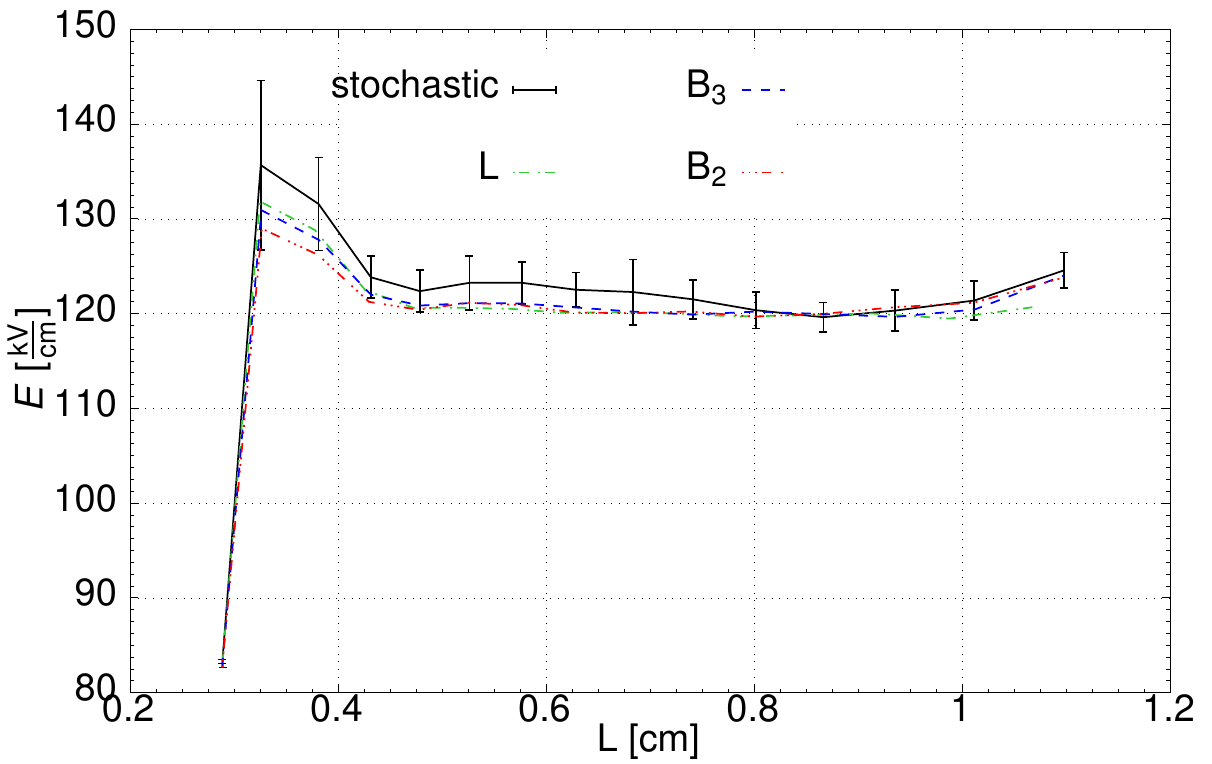}
    \includegraphics[width=\linewidth]{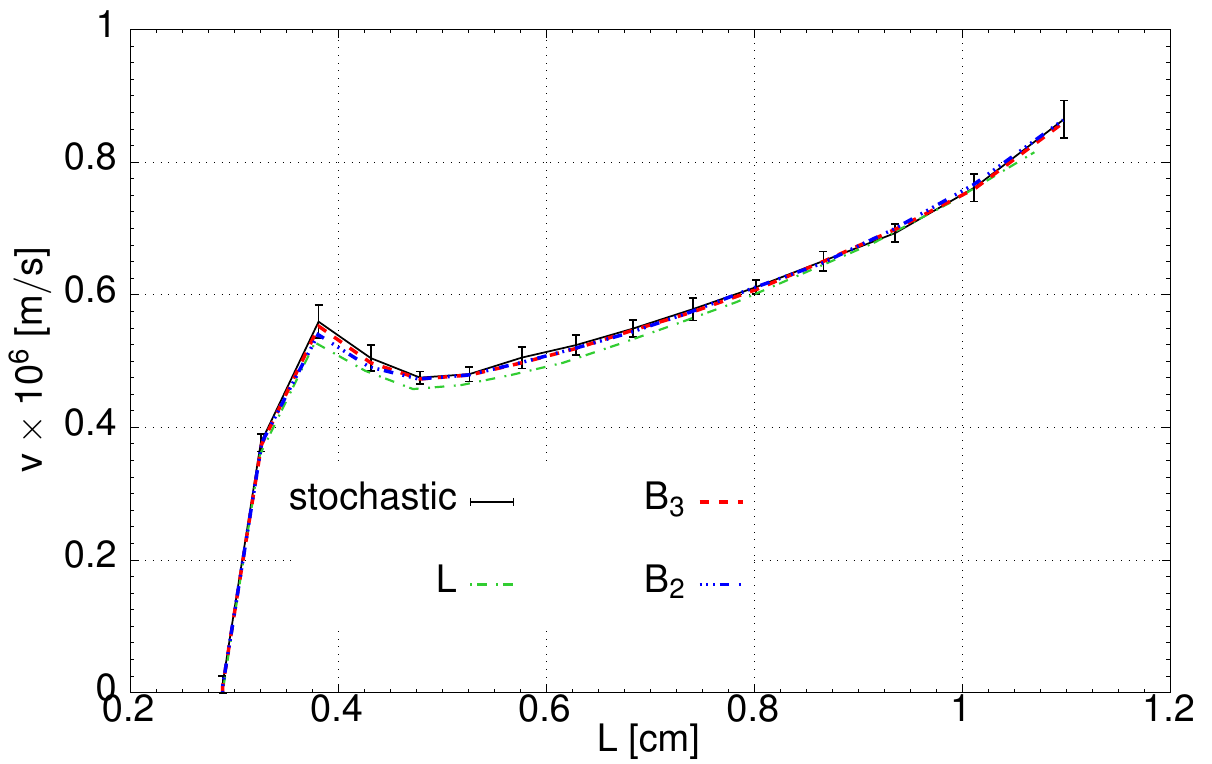}
    \caption{Top: $L(t)-vt$ ($v = 0.05$~cm/ns), where $L$ is the streamer length. Middle: maximal electric field versus streamer
      length. Bottom: velocity versus streamer
      length. L, B$_2$, and B$_3$ indicate the results with continuum photoionization. The average results for the ten stochastic runs are indicated with ``stochastic". The error bars indicate plus
and minus one standard deviation of the underlying samples.}
    \label{fig:3D-diff-rnds-constdx-10um-4}%
  \end{center}
\end{figure}

To more quantitatively compare the simulations, the streamer length $L$, maximal
electric field $E_\mathrm{max}$ and velocity $v$ are shown in figure
\ref{fig:3D-diff-rnds-constdx-10um-4}. The streamer length is defined as
$L(t) = L_\mathrm{domain} - z_{\mathrm{max}}(t)$, where $z_{\mathrm{max}}$ is
the $z$-coordinate where the electric field is maximal and
$L_\mathrm{domain}=1.25$~cm is the domain size. To show variations in length
more clearly, figure \ref{fig:3D-diff-rnds-constdx-10um-4} shows $L(t)-vt$ with
$v=0.05$~cm/ns, and $E_\mathrm{max}$ and $v$ are shown as a function of streamer
length, to allow for a comparison of streamer properties at the same propagation
length. The averages of the ten runs with stochastic photoionization are
indicated with error bars. The error bars indicate plus and minus one standard
deviation $\sigma$ of the underlying ten samples, and thus not the standard
error of the mean, which would be given by $\sigma/\sqrt{10}$.

For the streamer length, figure \ref{fig:3D-diff-rnds-constdx-10um-4} shows very
good agreement between the averaged stochastic results and the continuum case
with Bourdon's parameters ($B_2$ and $B_3$). With Luque's photoionization
parameters ($L$) the streamer is a bit slower, but the difference is less than
$0.5 \, \textrm{mm}$ at any time. This is probably related to the slightly
different definition of Luque's parameters, as discussed in section
\ref{sec:photoi-continuum-approach}.


The maximal electric field first increases rapidly, indicating that a streamer
has formed, and it then decreases slowly as the streamer propagates across the
gap. When the streamer approaches the bottom boundary an increase is again
visible, due to the compression of a voltage difference in a small region. The
standard deviation of the maximal field is initially around $10$~kV/cm in the
stochastic runs, but it becomes considerably smaller at later times. This is due
to the larger fluctuations in the early stage of the discharge, as also
illustrated in figure~\ref{fig:3D-diff-rnds-constdx-10um-photo-5}. These
fluctuations are probably also the reason why the average maximal field is a bit
higher than in the continuum results.

Fluctuations in the streamer velocity are much smaller than in the electric
field, and all photoionization approaches show good agreement. Overall, we
conclude that under the conditions used in this paper, stochastic photoionization leads
to relatively small fluctuations in streamer properties, and that we find good
agreement between the average stochastic behavior and the continuum results.

\subsection{Comparison with axisymmetric simulations}
\label{sec:Cylindircal}

Single streamers have often been simulated with axisymmetric
models~\cite{bagheri_comparison_2018,luque_positive_2008,aleksandrov_simulation_1996,vitello_simulation_1994,dhali_twodimensional_1987,bourdon_influence_2010},
which are computationally much cheaper than 3D models. When stochastic effects
are included, there are a couple of important differences between 3D and
axisymmetric models. With axial symmetry, streamers cannot deviate off-axis,
unless they branch into an unphysical conical shape~\cite{luque_density_2012}.
Furthermore, stochastic fluctuations are averaged over the azimuthal direction,
which leads to smaller fluctuations at larger radii (because they are averaged
over a larger volume).

\begin{figure}
  \begin{center}
    \includegraphics[width=1.0\linewidth]{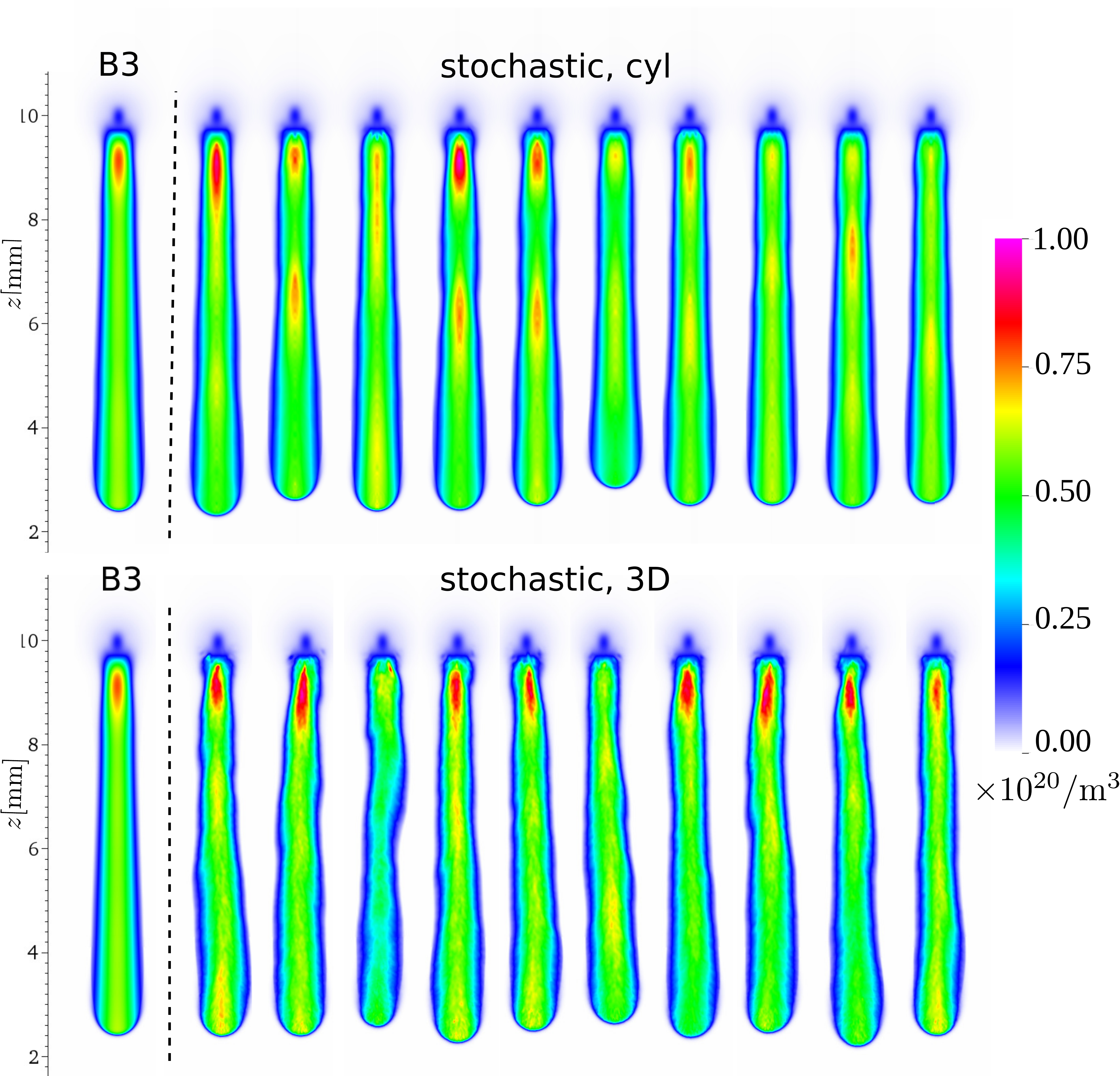}
    \caption{The electron density at $t=13$~ns for the axisymmetric simulations
      (top), and the 3D simulations (bottom). Cross sections of the electron
      density in the xz plane are shown for the 3D simulations. The ten
      stochastic runs for each case as well as the results with continuum
      photoionization (B$_3$) are shown. }
    \label{fig:2D-electron-density-ten-cases}
  \end{center}
\end{figure}

To investigate how axisymmetric simulations are affected by stochastic
photoionization, we repeat the simulations presented in section
\ref{sec:3D} with an axisymmetric model. The same procedure for stochastic
photoionization is used as in 3D, but the ionizing photons are now mapped to an
$(r,z)$ mesh, where $r^2 = x^2 + y^2$. Another difference is that the
cylindrical domain has a radius $R = 1.25 \, \textrm{cm}$, whereas the 3D
simulations were performed in a cube of $(1.25 \, \textrm{cm})^3$. This should
only have a small effect, because the streamer is quite far from the lateral
boundaries in both cases.

As before, ten runs are performed with different random numbers.
Figure~\ref{fig:2D-electron-density-ten-cases} shows the electron density at
$t=13$~ns for these ten runs, together with cross sections of the 3D simulations
for comparison. In all cases the axisymmetric streamer bridges the gap without
branching. Fluctuations in propagation length and in electron density are
comparable to those in the 3D simulations. Lateral deviations are of course only
visible in the 3D results.

\begin{figure}%
  \begin{center}
   \includegraphics[width=\linewidth]{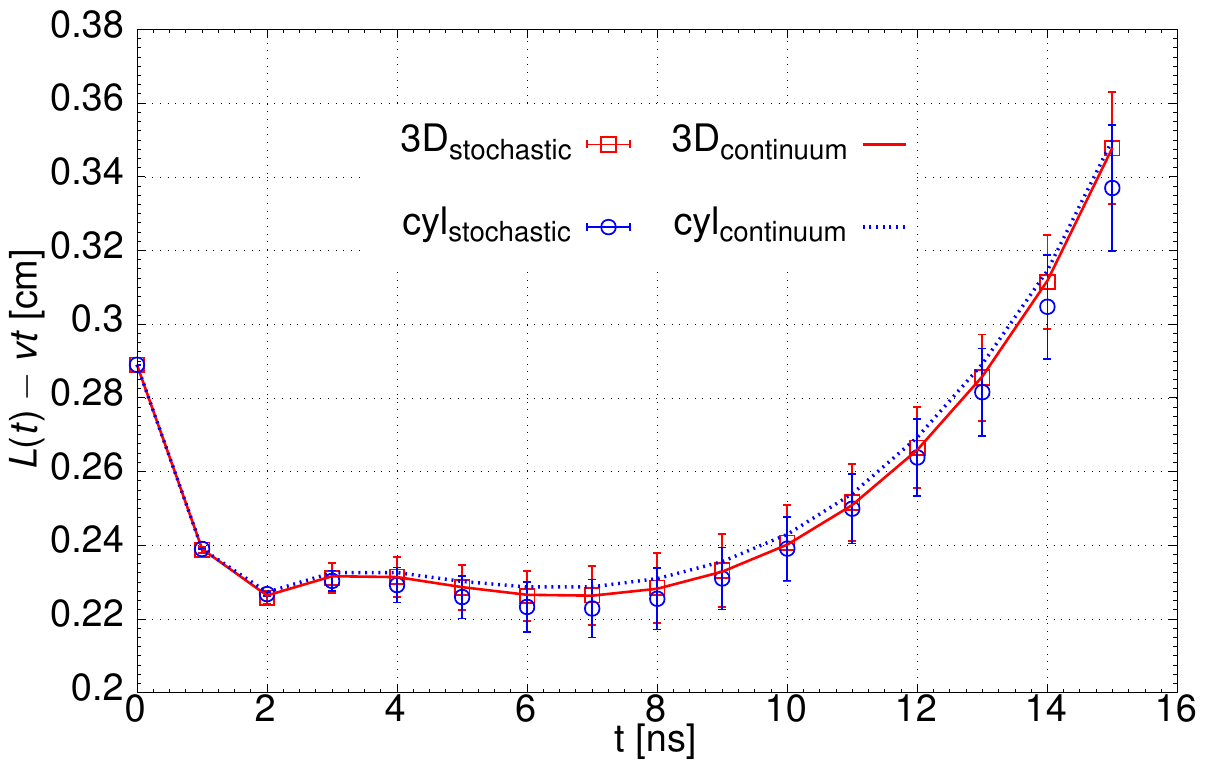}
    \includegraphics[width=\linewidth]{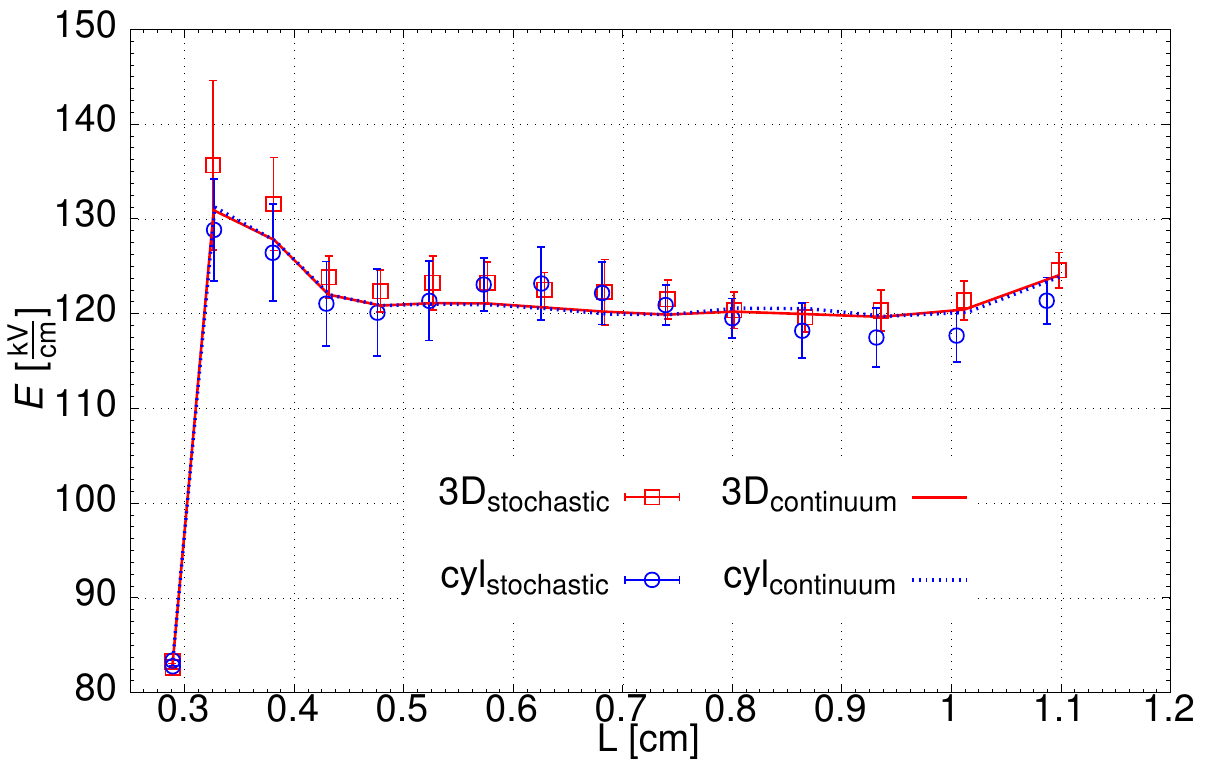}
    \includegraphics[width=\linewidth]{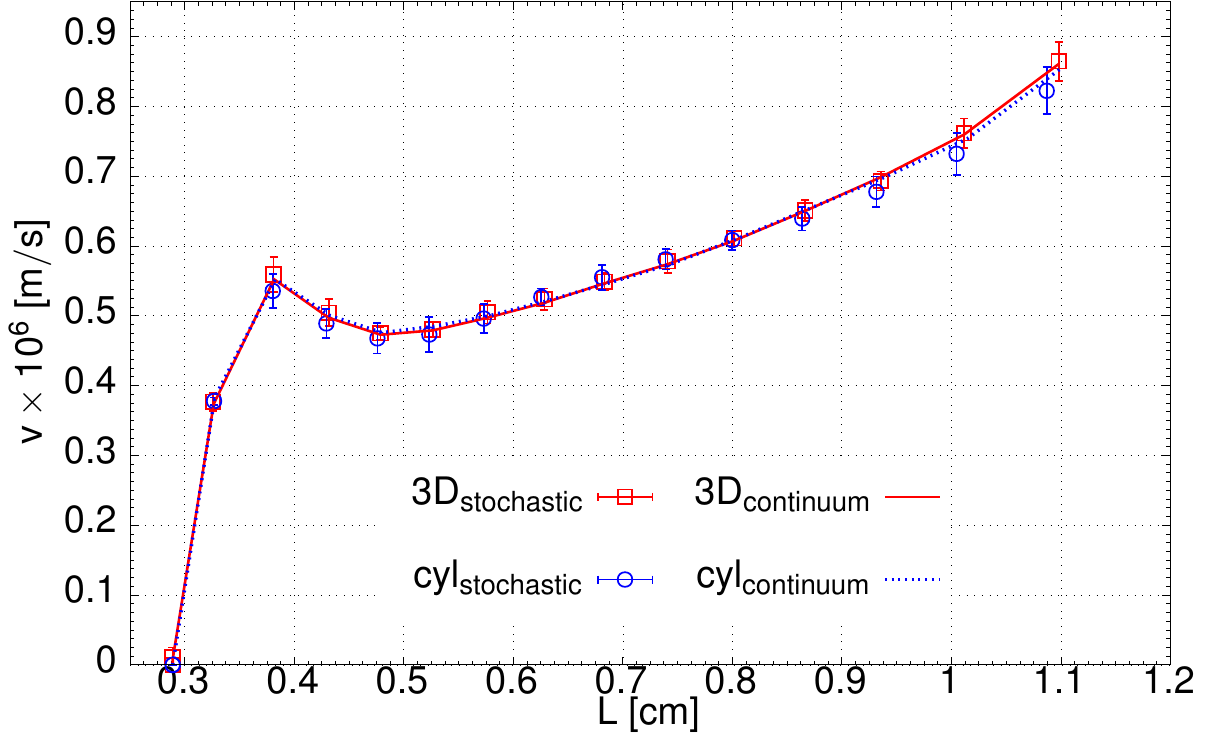}
    \caption{3D vs. axisymmetric. From top to bottom: $L(t)-v t$ ($v = 0.05$~cm/ns), maximal electric field, velocity. Results with continuum (B$_3$) and stochastic photoionization are shown. The axisymmetric results are indicated with ``cyl", and 3D results with ``3D".}
    \label{fig:3Dvscyl-analysis}%
  \end{center}
\end{figure}

For a more quantitative comparison, figure~\ref{fig:3Dvscyl-analysis} shows
streamer length, maximal electric field and velocity (as in section
\ref{sec:3D}). With stochastic photoionization, there is reasonably good
agreement between the 3D and axisymmetric results. The error bars indicate plus
and minus one standard deviation of the underlying samples, as in
section~\ref{sec:3D}. Differences up to about one standard deviation are visible
in the maximal electric field, and on average the maximal field is a bit higher
in the 3D simulations. This is probably due to small deformations of the
streamer's shape that can enhance the field in 3D, see e.g. figure
\ref{fig:E-ne-time-3D}.

Results with continuum photoionization are also included in figure
\ref{fig:3Dvscyl-analysis}, using Bourdon's three-term parameters. Without
stochastic effects, differences can only arise due to the different numerical
discretization of the equations, and due to the slightly different computational
domain. The agreement between the 3D and axisymmetric results is excellent, in
particular in terms of the maximal field and the streamer's velocity. The
agreement is also clear from figure \ref{fig:2D-electron-density-ten-cases},
which shows the electron density at $t =13 \, \textrm{ns}$ for both cases (on
the left).

\begin{figure}
\begin{center}
 \includegraphics[width=\linewidth]{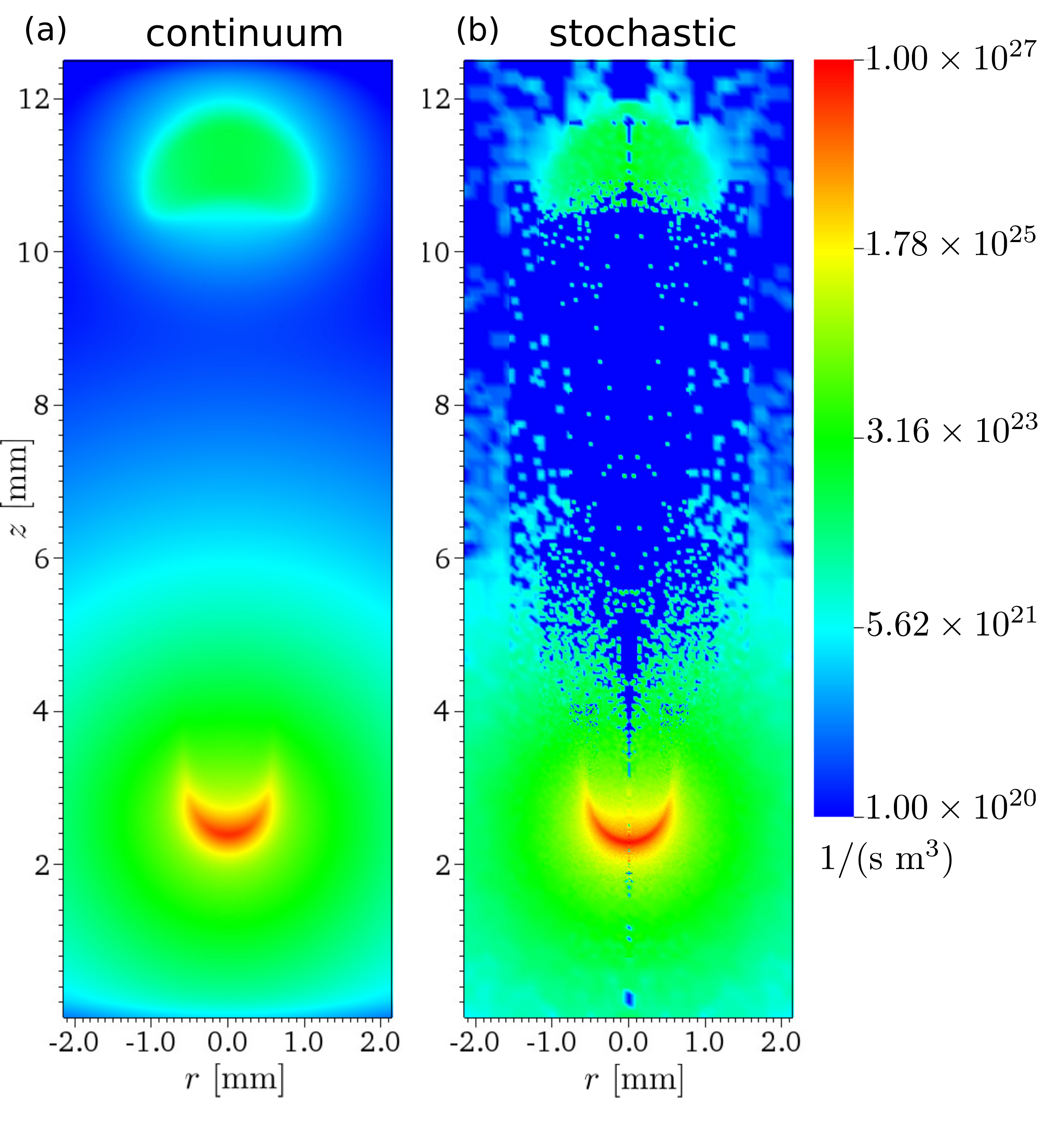}
 \caption{The production rate of photoelectrons (on a logarithmic scale) at
   t=13~ns in an axisymmetric model using (a) continuum photoionization ($B_3$)
   and (b) stochastic photoionization.}
 \label{fig:photo-profile-2D-Cyl}
\end{center}
\end{figure}

As mentioned above, stochastic noise is smaller at larger radii in axisymmetric
simulations, because the grid cells there correspond to larger volumes. This is
illustrated in figure~\ref{fig:photo-profile-2D-Cyl}, which shows the stochastic
and continuum production rate of photoelectrons at $t=13$~ns on a logarithmic
scale. Near the axis, the fluctuations are clearly strongest. Note that the
continuum and stochastic profile agree reasonably well, at least on a
logarithmic scale. We remark that with the stochastic approach the number of
ionizing photons produced during a time step depends on the length of the time
step (see section~\ref{sec:photoi-stochastic-approach}), which was here about
$2 \, \textrm{ps}$.

In conclusion, stochastic fluctuations also lead to small fluctuations in
streamer properties when an axisymmetric model is used. The fluctuations are
comparable to the ones observed in 3D, but lateral deviations in the streamer's
direction can no longer be modeled. Overall, the results are in good agreement,
both between axisymmetric and 3D and between continuum and stochastic
photoionization. This agreement will of course only hold as long as the 3D
streamers are approximately axisymmetric, which is the case here.

\subsection{Changing the amount of photoionization}
\label{sec:Photo-electron reduction}

\begin{figure}%
  \begin{center}
    \includegraphics[width=1.0\linewidth]{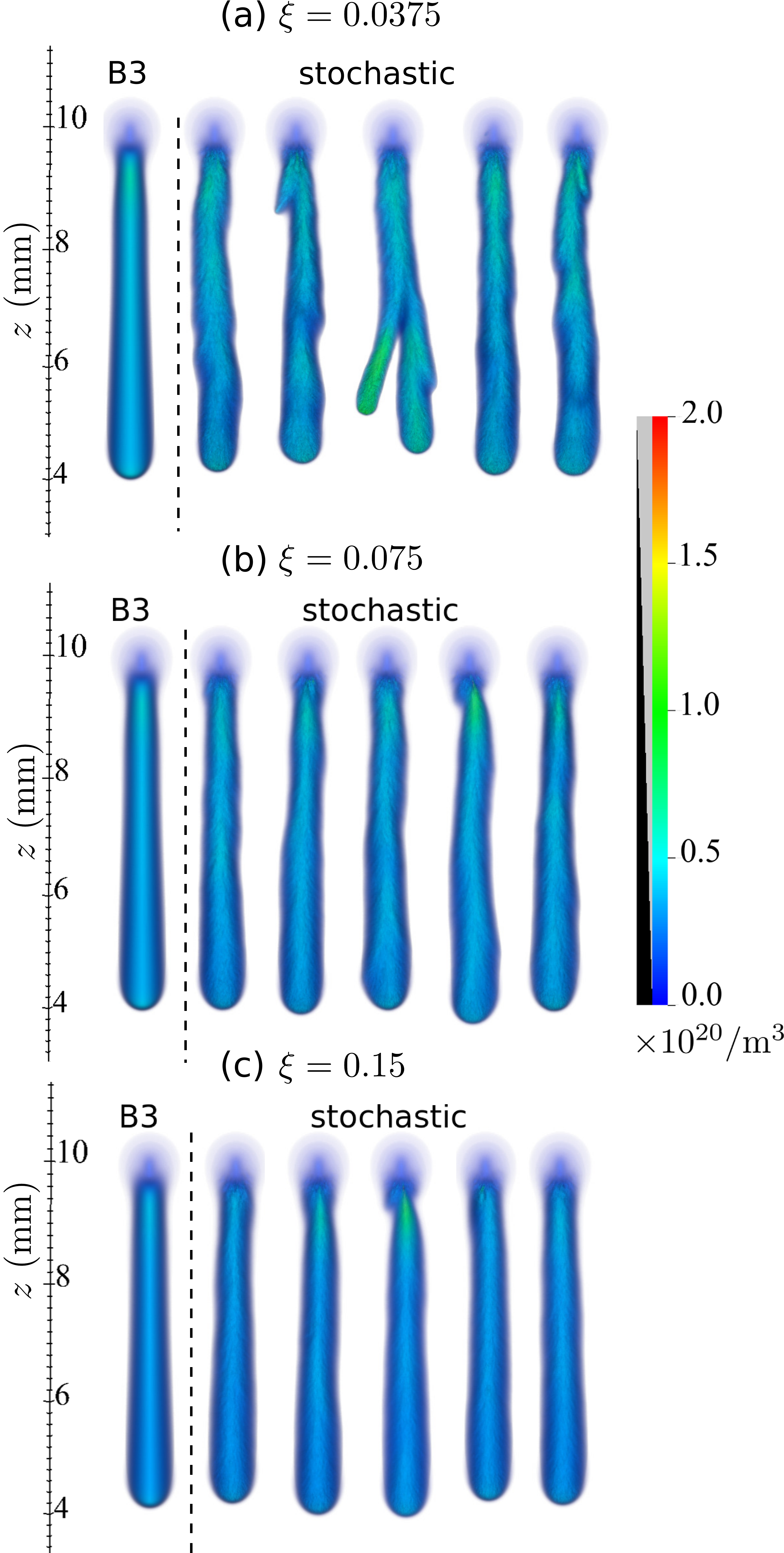}
    \caption{Volume renders of electron density at $t=12$~ns for 3D simulations
      with (a) $\xi=0.0375$, (b) $\xi=0.075$, (c) $\xi=0.15$. The parameter
      $\xi$ controls the amount of photoionization, see equation
      (\ref{equ:Photo-source-term}). Five stochastic runs, each with a different
      initial state of the random number generator are shown on the right. The
      results with continuum photoionization (B$_3$) are shown on the left.}
    \label{fig:3D-diff-eta-photo}%
  \end{center}
\end{figure}

In this section we investigate how changing the amount of photoionization
affects the streamer's propagation. We adjust the amount of photoionization
by changing the parameter $\xi$ in equation (\ref{equ:Photo-source-term}). Other
ways to vary the amount of photoionization are changing the gas mixture or the
gas pressure, which affects the generation rate as well as the absorption length 
of the UV photons.

Using the conditions from section \ref{sec:3D}, in which $\xi$ was set to
$0.075$, we have performed two additional sets of simulations: five runs for
$\xi = 0.0375$ and five runs for $\xi = 0.15$. These cases correspond to half
and double the amount of photoionization, although only approximately, since the
discharge itself will also change with $\xi$. Figure~\ref{fig:3D-diff-eta-photo}
shows volume renders of the electron density for these simulations at $t=12$~ns,
together with results from section \ref{sec:3D}. For comparison, results with
continuum photoionization are also shown in figure~\ref{fig:3D-diff-eta-photo},
using Bourdon's three-term parameters.

Whereas the results with continuum photoionization hardly change, the stochastic
simulations are surprisingly sensitive 
to the amount of photoionization. For
$\xi = 0.0375$, fluctuations in the electron density profile are significantly
larger than in the other cases. In one of the five runs, these fluctuations lead
to branching, and the branched streamer has the shortest length. For
$\xi = 0.075$ and $\xi = 0.15$, the streamers always bridge the gap without
branching. The increase in photoionization for $\xi = 0.15$ leads to a smoother
electron density profile than for $\xi = 0.075$, but differences in propagation
length appear to be similar. Figure~\ref{fig:3D-diff-eta-photo} also shows that
streamer velocities are not sensitive to the amount of photoionization, except
for the case with streamer branching, which is in agreement with the results of
\cite{wormeester_probing_2010}.

In conclusion, the simulations with stochastic photoionization are much more
sensitive to the amount of photoionization than those with continuum
photoionization. With stochastic photoionization, a decrease in the number of
photoelectrons initially leads to more noise, which causes different types of
protrusion to form in the streamer channel. If the fluctuations are strong
enough, these protrusions can even lead to spontaneous branching. This suggests
that stochastic photoionization should be taken into account to model streamer
branching, in particular in gas mixtures with less photoionization than air. If
we increase the amount of photoelectrons, the streamer channel develops more
smoothly than in section \ref{sec:3D}, but there are still small differences in
the propagation length.



\section{Conclusions}
\label{sec:conclusions}

For positive streamer discharges, photoionization is often an important
mechanism to provide free electrons ahead of the streamer. Depending on the
number of photoionization events and on the volume in which they take place, the
photoionization profile can vary from smooth to noisy. In this paper, we have
investigated how important such stochastic fluctuations are for positive
streamers in atmospheric air. We remark that photoionization is not the only
possible source of stochastic fluctuations. Spatial variations in the background
ionization density can also affect the propagation and branching of streamers,
see e.g.~\cite{teunissen_simulating_2017}. Conversely, in conditions where a
reasonably high background ionization is included, the stochastic effects of
photoionization do not play a significant role.

We performed numerical simulations with a 3D plasma fluid model in which
photoionization could be included as a stochastic or as a continuum process.
Stochastic photoionization profiles were computed with a Monte Carlo method, in
which individual ionizing photons were modeled. Continuum photoionization
profiles were computed using the Helmholtz approximation. The evolution of
positive streamers between two planar electrodes was simulated, in a background
field of $15 \, \textrm{kV/cm}$ which is about half of the breakdown field. With
stochastic photoionization, we observed fluctuations in streamer properties such
as maximal electric field, velocity and electron density. In the test cases
considered here, these fluctuations were not strong enough to cause branching.
When the stochastic results were averaged, they were in good agreement with the
results obtained with continuum photoionization.

Axisymmetric models are commonly used for the simulation of single streamers,
because they are computationally much cheaper. However, due to their imposed
symmetry, such models cannot properly capture stochastic fluctuations. To see
how much this affects simulation results, we compared our 3D simulations to
axisymmetric ones. With continuum photoionization, the results agreed very well.
With stochastic photoionization, the axisymmetric results showed similar
fluctuations in streamer length, maximal electric field and streamer velocity as
the 3D simulations. On average, these streamer properties were in good agreement
with the 3D simulations.

Finally, we compared stochastic and continuum photoionization for cases with
half and double the amount of photoionization. With double the amount of
photoionization, stochastic fluctuations were reduced, and there was good
agreement between the stochastic and continuum results. With half the amount of
photoionization, stochastic fluctuations became much more important, and
branching started to occur. Our results are therefore surprisingly sensitive to
the amount of photoionization. Because of this sensitivity, we expect that
stochastic photoionization can cause streamer branching in other discharge
configurations in atmospheric air (i.e., without an artificially reduced
photoionization level). For example, the background field used here is
homogeneous, whereas the field from a pointed electrode can have strong
lateral components that accelerate branching. Furthermore, for discharges
developing in lower background fields the stochastic effects of photoionization
can be stronger.

In conclusion, we find good agreement between the averaged stochastic
simulations and the continuum simulations in our test cases in atmospheric air.
However, stochastic fluctuations significantly increase when the amount of
photoionization is reduced. In other gas mixtures or different discharge
conditions, the stochastic effects of photoionization could therefore play an
important role in the propagation and branching of positive streamers. 


\section{Acknowledgments}
B.B. acknowledges funding through the Dutch TTW-project 15052, and J.T. through postdoctoral fellowship 12Q6117N of the Belgian-Flemish FWO. The authors gratefully acknowledge Ute Ebert for her profound comments and discussions during this work.

The simulation code used in this paper is available at \url{https://gitlab.com/MD-CWI-NL/afivo-streamer}. Moreover the input files for generating the results together with the output files are provided in \url{https://doi.org/10.17026/dans-zec-672a}.

\section*{References}
\bibliography{jannis_bibtex_new}

\providecommand{\newblock}{}
\begin{thebibliography}{10}
\expandafter\ifx\csname url\endcsname\relax
  \def\url#1{{\tt #1}}\fi
\expandafter\ifx\csname urlprefix\endcsname\relax\def\urlprefix{URL }\fi
\providecommand{\eprint}[2][]{\url{#2}}

\bibitem{vitello_simulation_1994}
Vitello P~A, Penetrante B~M and Bardsley J~N 1994 {\em Physical Review E\/}
  {\bf 49} 5574--5598 ISSN 1095-3787
  \urlprefix\url{http://dx.doi.org/10.1103/PhysRevE.49.5574}

\bibitem{yi_experimental_2002}
Yi W~J and Williams P~F 2002 {\em J. Phys. D: Appl. Phys.\/} {\bf 35} 205--218
  ISSN 1361-6463 \urlprefix\url{http://dx.doi.org/10.1088/0022-3727/35/3/308}

\bibitem{ebert_review_2010}
Ebert U, Nijdam S, Li C, Luque A, Briels T and van Veldhuizen E 2010 {\em
  Journal of Geophysical Research\/} {\bf 115} ISSN 0148-0227
  \urlprefix\url{http://dx.doi.org/10.1029/2009JA014867}

\bibitem{nijdam_probing_2010}
Nijdam S, van~de Wetering F~M~J~H, Blanc R, van Veldhuizen E~M and Ebert U 2010
  {\em J. Phys. D: Appl. Phys.\/} {\bf 43} 145204 ISSN 1361-6463
  \urlprefix\url{http://dx.doi.org/10.1088/0022-3727/43/14/145204}

\bibitem{An_2007}
An W, Baumung K and Bluhm H 2007 {\em Journal of Applied Physics\/} {\bf 101}
  053302 ISSN 1089-7550 \urlprefix\url{http://dx.doi.org/10.1063/1.2437675}

\bibitem{sentman_red_1995}
Sentman D~D and Wescott E~M 1995 {\em Phys. Plasmas\/} {\bf 2} 2514 ISSN
  1070-664X \urlprefix\url{http://dx.doi.org/10.1063/1.871213}

\bibitem{Fridman_2005}
Fridman A, Chirokov A and Gutsol A 2005 {\em Journal of Physics D: Applied
  Physics\/} {\bf 38} R1–R24 ISSN 1361-6463
  \urlprefix\url{http://dx.doi.org/10.1088/0022-3727/38/2/R01}

\bibitem{Adamovich_2017}
Adamovich I, Baalrud S~D, Bogaerts A, Bruggeman P~J, Cappelli M, Colombo V,
  Czarnetzki U, Ebert U, Eden J~G, Favia P and et~al 2017 {\em Journal of
  Physics D: Applied Physics\/} {\bf 50} 323001 ISSN 1361-6463
  \urlprefix\url{http://dx.doi.org/10.1088/1361-6463/aa76f5}

\bibitem{kanazawa_observation_2011}
Kanazawa S, Kawano H, Watanabe S, Furuki T, Akamine S, Ichiki R, Ohkubo T,
  Kocik M and Mizeraczyk J 2011 {\em Plasma Sources Science and Technology\/}
  {\bf 20} 034010 ISSN 1361-6595
  \urlprefix\url{http://dx.doi.org/10.1088/0963-0252/20/3/034010}

\bibitem{starikovskaia_plasma-assisted_2014}
Starikovskaia S~M 2014 {\em J. Phys. D: Appl. Phys.\/} {\bf 47} 353001 ISSN
  1361-6463 \urlprefix\url{http://dx.doi.org/10.1088/0022-3727/47/35/353001}

\bibitem{nozaki_non-thermal_2013}
Nozaki T and Okazaki K 2013 {\em Catalysis Today\/} {\bf 211} 29--38 ISSN
  0920-5861 \urlprefix\url{http://dx.doi.org/10.1016/j.cattod.2013.04.002}

\bibitem{briels_positive_2008}
Briels T~M~P, Kos J, Winands G~J~J, van Veldhuizen E~M and Ebert U 2008 {\em J.
  Phys. D: Appl. Phys.\/} {\bf 41} 234004 ISSN 1361-6463
  \urlprefix\url{http://dx.doi.org/10.1088/0022-3727/41/23/234004}

\bibitem{nijdam_role_2016}
Nijdam S, Teunissen J, Takahashi E and Ebert U 2016 {\em Plasma Sources Science
  and Technology\/} {\bf 25} 044001 ISSN 1361-6595
  \urlprefix\url{http://dx.doi.org/10.1088/0963-0252/25/4/044001}

\bibitem{pancheshnyi_role_2005}
Pancheshnyi S 2005 {\em Plasma Sources Sci. Technol.\/} {\bf 14} 645--653 ISSN
  1361-6595 \urlprefix\url{http://dx.doi.org/10.1088/0963-0252/14/4/002}

\bibitem{zheleznyak_photoionization_1982}
Zheleznyak M~B, Mnatsakanian A~K and Sizykh S~V 1982 {\em Teplofizika Vysokikh
  Temperatur\/} {\bf 20} 423--428

\bibitem{pancheshnyi_photoionization_2014}
Pancheshnyi S 2014 {\em Plasma Sources Sci. Technol.\/} {\bf 24} 015023 ISSN
  1361-6595 \urlprefix\url{http://dx.doi.org/10.1088/0963-0252/24/1/015023}

\bibitem{Stephens_2016}
Stephens J, Fierro A, Beeson S, Laity G, Trienekens D, Joshi R~P, Dickens J and
  Neuber A 2016 {\em Plasma Sources Science and Technology\/} {\bf 25} 025024
  ISSN 1361-6595
  \urlprefix\url{http://dx.doi.org/10.1088/0963-0252/25/2/025024}

\bibitem{Stephens_2018}
Stephens J, Abide M, Fierro A and Neuber A 2018 {\em Plasma Sources Science and
  Technology\/} {\bf 27} 075007 ISSN 1361-6595
  \urlprefix\url{http://dx.doi.org/10.1088/1361-6595/aacc91}

\bibitem{Jiang_2018}
Jiang M, Li Y, Wang H, Zhong P and Liu C 2018 {\em Physics of Plasmas\/} {\bf
  25} 012127 ISSN 1089-7674 \urlprefix\url{http://dx.doi.org/10.1063/1.5019478}

\bibitem{luque_photoionization_2007}
Luque A, Ebert U, Montijn C and Hundsdorfer W 2007 {\em Appl. Phys. Lett.\/}
  {\bf 90} 081501 ISSN 0003-6951
  \urlprefix\url{http://dx.doi.org/10.1063/1.2435934}

\bibitem{bourdon_efficient_2007}
Bourdon A, Pasko V~P, Liu N~Y, Célestin S, Ségur P and Marode E 2007 {\em
  Plasma Sources Sci. Technol.\/} {\bf 16} 656--678 ISSN 1361-6595
  \urlprefix\url{http://dx.doi.org/10.1088/0963-0252/16/3/026}

\bibitem{chanrion_pic-mcc_2008}
Chanrion O and Neubert T 2008 {\em Journal of Computational Physics\/} {\bf
  227} 7222--7245 ISSN 0021-9991
  \urlprefix\url{http://dx.doi.org/10.1016/j.jcp.2008.04.016}

\bibitem{bagheri_comparison_2018}
Bagheri B, Teunissen J, Ebert U, Becker M~M, Chen S, Ducasse O, Eichwald O,
  Loffhagen D, Luque A, Mihailova D, Plewa J~M, van Dijk J and Yousfi M 2018
  {\em Submitted to Plasma Sources Sci. Technol.\/}

\bibitem{Xiong_2014}
Xiong Z and Kushner M~J 2014 {\em Plasma Sources Science and Technology\/} {\bf
  23} 065041 ISSN 1361-6595
  \urlprefix\url{http://dx.doi.org/10.1088/0963-0252/23/6/065041}

\bibitem{Luque_2011}
Luque A and Ebert U 2011 {\em Physical Review E\/} {\bf 84} ISSN 1550-2376
  \urlprefix\url{http://dx.doi.org/10.1103/PhysRevE.84.046411}

\bibitem{li_comparison_2012}
Li C, Teunissen J, Nool M, Hundsdorfer W and Ebert U 2012 {\em Plasma Sources
  Science and Technology\/} {\bf 21} 055019 ISSN 1361-6595
  \urlprefix\url{http://dx.doi.org/10.1088/0963-0252/21/5/055019}

\bibitem{Teunissen_2016}
Teunissen J and Ebert U 2016 {\em Plasma Sources Science and Technology\/} {\bf
  25} 044005 ISSN 1361-6595
  \urlprefix\url{http://dx.doi.org/10.1088/0963-0252/25/4/044005}

\bibitem{teunissen_simulating_2017}
Teunissen J and Ebert U 2017 {\em Journal of Physics D: Applied Physics\/} {\bf
  50} 474001 ISSN 1361-6463
  \urlprefix\url{http://dx.doi.org/10.1088/1361-6463/aa8faf}

\bibitem{dutton_survey_1975}
Dutton J 1975 {\em Journal of Physical and Chemical Reference Data\/} {\bf 4}
  577--856 \urlprefix\url{http://dx.doi.org/doi/10.1063/1.555525}

\bibitem{hartmann_theoretical_1984}
Hartmann G 1984 {\em IEEE Transactions on Industry Applications\/} {\bf IA-20}
  1647--1651 ISSN 0093-9994, 1939-9367
  \urlprefix\url{http://ieeexplore.ieee.org/document/4504655/}

\bibitem{hagelaar_solving_2005}
Hagelaar G~J~M and Pitchford L~C 2005 {\em Plasma Sources Science and
  Technology\/} {\bf 14} 722--733 ISSN 1361-6595
  \urlprefix\url{http://dx.doi.org/10.1088/0963-0252/14/4/011}

\bibitem{Teunissen_afivo_2018}
Teunissen J and Ebert U 2018 {\em Computer Physics Communications\/} {\bf 233}
  156–166 ISSN 0010-4655
  \urlprefix\url{http://dx.doi.org/10.1016/j.cpc.2018.06.018}

\bibitem{teunissen_3d_2015}
Teunissen J 2015 {\em 3D {Simulations} and {Analysis} of {Pulsed}
  {Discharges}\/} {PhD} {Thesis} Technische Universiteit Eindhoven,
  http://repository.tue.nl/801516

\bibitem{luque_positive_2008}
Luque A, Ratushnaya V and Ebert U 2008 {\em J. Phys. D: Appl. Phys.\/} {\bf 41}
  234005 ISSN 1361-6463
  \urlprefix\url{http://dx.doi.org/10.1088/0022-3727/41/23/234005}

\bibitem{aleksandrov_simulation_1996}
Aleksandrov N~L and Bazelyan E~M 1996 {\em J. Phys. D: Appl. Phys.\/} {\bf 29}
  740--752 ISSN 1361-6463
  \urlprefix\url{http://dx.doi.org/10.1088/0022-3727/29/3/035}

\bibitem{dhali_twodimensional_1987}
Dhali S~K and Williams P~F 1987 {\em Journal of Applied Physics\/} {\bf 62}
  4696--4707 ISSN 1089-7550 \urlprefix\url{http://dx.doi.org/10.1063/1.339020}

\bibitem{bourdon_influence_2010}
Bourdon A, Bonaventura Z and Celestin S 2010 {\em Plasma Sources Sci.
  Technol.\/} {\bf 19} 034012 ISSN 1361-6595
  \urlprefix\url{http://dx.doi.org/10.1088/0963-0252/19/3/034012}

\bibitem{luque_density_2012}
Luque A and Ebert U 2012 {\em Journal of Computational Physics\/} {\bf 231}
  904--918 ISSN 0021-9991
  \urlprefix\url{http://dx.doi.org/10.1016/j.jcp.2011.04.019}

\bibitem{wormeester_probing_2010}
Wormeester G, Pancheshnyi S, Luque A, Nijdam S and Ebert U 2010 {\em J. Phys.
  D: Appl. Phys.\/} {\bf 43} 505201 ISSN 1361-6463
  \urlprefix\url{http://dx.doi.org/10.1088/0022-3727/43/50/505201}

\end{thebibliography}

\end{document}